\documentclass[aps,preprint,nofootinbib]{revtex4}%
\usepackage{amsfonts}
\usepackage{amsmath}
\usepackage{amssymb}
\usepackage{graphicx}%
\providecommand{\U}[1]{\protect\rule{.1in}{.1in}}
\begin{document}
\title[Short title for running header]{Analysis of the Hamiltonian formulations of linearized General Relativity}
\author{K. R. Green}
\email{Kevin.Green@uoit.ca}
\affiliation{Faculty of Science, University of Ontario Institute of Technology, L1H 7K4,
Oshawa, Canada}
\author{N. Kiriushcheva}
\email{nkiriush@uwo.ca}
\affiliation{Faculty of Arts and Social Science, Huron University College, N6G 1H3 and
Department of Applied Mathematics, University of Western Ontario, N6A 5B7,
London, Canada}
\author{S. V. Kuzmin}
\email{skuzmin@uwo.ca}
\affiliation{Faculty of Arts and Social Science, Huron University College, N6G 1H3 and
Department of Applied Mathematics, University of Western Ontario, N6A 5B7,
London, Canada}
\pacs{PACS number}

\begin{abstract}
The different forms of the Hamiltonian formulations of linearized General
Relativity/spin-2 theories are discussed in order to show their similarities
and differences. It is demonstrated that in the linear model, non-covariant
modifications to the initial covariant Lagrangian (similar to those
modifications used in full gravity) are in fact unnecessary. The Hamiltonians
and the constraints are different in these two formulations but the structure
of the constraint algebra and the gauge invariance derived from it are the
same. It is shown that these equivalent Hamiltonian formulations are related
to each other by a canonical transformation which is explicitly given. The
relevance of these results to the full theory of General Relativity is briefly discussed.

\end{abstract}
\date{\today}
\maketitle
\eid{ }
\accepted{}

%\date[Date text]{date}
%\volumeyear{year}
%\volumenumber{number}
%\issuenumber{number}
%\eid{identifier}
%\date{}
%\date[Date text]{date}
%\received[Received text]{date}

%\revised[Revised text]{date}

%\accepted[Accepted text]{date}

%\published[Published text]{date}

\thispagestyle{empty}

%\startpage{101}
%\endpage{102}
%\tableofcontents

\section{Introduction}

Linearized Gravity is discussed in all books on General Relativity (GR) and
its Hamiltonian analysis is the subject of many articles, for example
\cite{Baaklini, Blas, McKeon, Pad, Ferrari, Ghalati, Leclerc-1, Leclerc-2,
Leclerc-3}. However, we want to explore this subject again for three reasons.
First of all, there are different Hamiltonian formulations of linearized GR
but the relation among them has not been analyzed. As well, there also exist
some points which are not discussed in the literature that should raise
questions from outsiders to the field. Clarifying these points is our first,
mainly pedagogical, goal. The second reason for our analysis is to understand
the role that the linear approximation plays as a guide for the analysis of
the full theory of GR. Finally, the third reason is to closely examine some
customs in the Hamiltonian formulation of full GR involving simplifications,
\textit{a priori} assumptions, and construction of expected results rather
than performing direct calculations on the unmodified model. Investigating the
linear approximation can perhaps provide insight on such assumptions and/or
constructions used in the full theory of GR.

The Hamiltonian formulation of linearized GR can be approached from two quite
different directions. One originates from the well-known relation of
linearized GR to spin-2 theory. For the purpose of this article we use term
\textquotedblleft spin-2\textquotedblright\ as a short name for the gauge
theory of the massless non-interacting symmetric second rank tensor field
\cite{Fierz-Pauli}. Another direction is to start from the Einstein-Hilbert
action and linearize it. Both approaches produce the same result if one
chooses certain values of parameters appearing in the most general spin-2
Lagrangian. In addition to these approaches, we also consider the modified
Lagrangian which is the linearized version of the Dirac Lagrangian
\cite{Dirac}. Actually, the latter Lagrangian is more popular in the
literature, however the need for such a modification as well as a comparison
with results obtained before any modification have not been analyzed.

\section{Lagrangian densities}

\subsection{Spin-2}

We start from the action of a spin-2 field $h_{\alpha\beta}=h_{\beta\alpha}$
that can be built out of scalars which are quadratic in the derivatives
$h_{\alpha\beta,\gamma}$. Such a method was proposed by Feynman, which can be
found in \cite{Feynman} (and also in Appendix A of \cite{Pad}). The analysis
shows that out of all possible combinations, there are only five distinct
terms which are quadratic in derivatives. This can be presented in the
following form:%

\begin{equation}
L=\frac{1}{4}h_{\alpha\beta,\gamma}h_{\mu\nu,\rho}\left[  c_{1}\eta
^{\alpha\beta}\eta^{\mu\nu}\eta^{\gamma\rho}+c_{2}\eta^{\alpha\mu}\eta
^{\beta\nu}\eta^{\gamma\rho}+c_{3}\eta^{\alpha\gamma}\eta^{\beta\nu}\eta
^{\mu\rho}+c_{4}\eta^{\alpha\rho}\eta^{\beta\nu}\eta^{\gamma\mu}+c_{5}%
\eta^{\alpha\beta}\eta^{\gamma\mu}\eta^{\nu\rho}\right]  , \label{eq.1}%
\end{equation}
where $\eta^{\alpha\beta}=diag(-+++...)$ is the Minkowski tensor. Greek
letters are used for spacetime indices ($\alpha=0,1,2,3,...$).

The requirement that the field equations of (\ref{eq.1}) should be invariant
under the following gauge transformation%

\begin{equation}
h_{\alpha\beta}\rightarrow h_{\alpha\beta}-\xi_{\alpha,\beta}-\xi
_{\beta,\alpha} \label{eq.2}%
\end{equation}
puts severe restrictions on the parameters $c_{i}$,%

\begin{equation}
c_{1}=-c_{2}=1,\left.  {}\right.  c_{3}+c_{4}=-c_{5}=2. \label{eq.3}%
\end{equation}

Consequently, except for an overall scaling factor of $\frac{1}{4}$, we have a
one parameter family of Lagrangians. Note that the difference between the two
terms in (\ref{eq.1}) proportional to $c_{3}$ and $c_{4}$ is a total
divergence in the Minkowski space, e.g.%

\begin{equation}
h_{\alpha\beta,\rho}h_{\mu\nu,\gamma}=h_{\alpha\beta,\gamma}h_{\mu\nu,\rho
}+\partial_{\rho}\left(  h_{\alpha\beta}h_{\mu\nu,\gamma}\right)
-\partial_{\gamma}\left(  h_{\alpha\beta}h_{\mu\nu,\rho}\right)  .
\label{eq.4}%
\end{equation}

\subsection{Linear approximation of GR}

We can start from the Einstein-Hilbert (EH) Lagrangian%

\begin{equation}
L_{EH}=\sqrt{-g}R. \label{eq.5}%
\end{equation}
Alternatively, by dropping a total divergence in the expression (\ref{eq.5}),
and only considering the \textquotedblleft gamma-gamma\textquotedblright\ part
of $L_{EH}$, one obtains \cite{Landau, Carmeli}%

\begin{equation}
L_{\Gamma\Gamma}=\sqrt{-g}g^{\alpha\beta}\left(  \Gamma_{\alpha\nu}^{\mu
}\Gamma_{\beta\mu}^{\nu}-\Gamma_{\alpha\beta}^{\nu}\Gamma_{\nu\mu}^{\mu
}\right)  =\frac{1}{4}\sqrt{-g}B^{\alpha\beta\gamma\mu\nu\rho}g_{\alpha
\beta,\gamma}g_{\mu\nu,\rho}~, \label{eq.6}%
\end{equation}
where%

\begin{equation}
B^{\alpha\beta\gamma\mu\nu\rho}=g^{\alpha\beta}g^{\gamma\rho}g^{\mu\nu
}-g^{\alpha\mu}g^{\beta\nu}g^{\gamma\rho}+2g^{\alpha\rho}g^{\beta\nu}%
g^{\gamma\mu}-2g^{\alpha\beta}g^{\gamma\mu}g^{\nu\rho}. \label{eq.7}%
\end{equation}

We perform linearization around Minkowski space: $g_{\alpha\beta}=\eta
_{\alpha\beta}+h_{\alpha\beta}$. To preserve the property of the metric
tensor: $g_{\alpha\beta}g^{\alpha\gamma}=\delta_{\beta}^{\gamma}$ in linear
approximation, the contravariant tensor $g^{\mu\nu}$ should be chosen as:
$g^{\mu\nu}=\eta^{\mu\nu}-h^{\mu\nu}$. Substituting the linear approximation
of the metric tensor into (\ref{eq.5}), keeping only terms quadratic in the
derivatives of $h_{\alpha\beta,\gamma}$, and disregarding a total divergence,
we obtain%

\begin{equation}
L_{lin}=\frac{1}{4}h_{\alpha\beta,\gamma}h_{\mu\nu,\rho}\left(  \eta
^{\alpha\beta}\eta^{\gamma\rho}\eta^{\mu\nu}-\eta^{\alpha\mu}\eta^{\beta\nu
}\eta^{\gamma\rho}+2\eta^{\alpha\rho}\eta^{\beta\nu}\eta^{\gamma\mu}%
-2\eta^{\alpha\beta}\eta^{\gamma\mu}\eta^{\nu\rho}\right)  . \label{eq.8}%
\end{equation}

Another approach to the linearization of GR, the most straightforward, is to
use $L_{\Gamma\Gamma}$ (\ref{eq.6}), applying the simple rules%

\begin{equation}
g^{\alpha\beta}\rightarrow\eta^{\alpha\beta},\quad\sqrt{-g}\rightarrow1,\quad
g_{\alpha\beta,\gamma}\rightarrow h_{\alpha\beta,\gamma}~, \label{eq.8a}%
\end{equation}
which immediately gives the Lagrangian of (\ref{eq.1}) with parameters
satisfying (\ref{eq.3}) and $c_{3}=0$. These terms directly follow from the EH
action and after linearization (in this order only) we can do an integration
by parts with two terms. The terms $c_{3}h_{\alpha\beta,\gamma}h_{\mu\nu,\rho
}\eta^{\alpha\gamma}\eta^{\beta\nu}\eta^{\mu\rho}$ \ \ and $c_{4}%
h_{\alpha\beta,\gamma}h_{\mu\nu,\rho}\eta^{\alpha\rho}\eta^{\beta\nu}\eta
^{\mu\gamma}$ are equivalent up to a surface term, as in \ref{eq.4}), but this
cannot be generalized to the corresponding terms in full GR, e.g. for
$c_{3}\sqrt{-g}g_{\alpha\beta,\gamma}g_{\mu\nu,\rho}g^{\alpha\gamma}%
g^{\beta\nu}g^{\mu\rho}$ and $c_{4}\sqrt{-g}g_{\alpha\beta,\gamma}g_{\mu
\nu,\rho}g^{\alpha\rho}g^{\beta\nu}g^{\mu\gamma}$.

\subsection{Modifications}

In the literature on the Hamiltonian formulation of linearized GR, one can
initial Lagrangians with $c_{3}=0$ (e.g. \cite{Ferrari}) or $c_{4}=0$ (e.g.
\cite{Leclerc-3}). It is perfectly correct to consider both cases for
spin-2/linear approximation of GR but it is not clear why it is necessary to
perform an additional integration by parts to set $c_{4}=0$\ since direct
linearization gives $c_{3}=0$. Moreover, the reason for requiring further
modification of the covariant Lagrangian and to present it in a non-covariant
form \textquotedblleft up to total space and time
derivatives\textquotedblright\ \cite{Baaklini} or make \textquotedblleft some
rearrangements\textquotedblright\ \cite{Ferrari} is not given. Of course, it
is not difficult to restore these integrations once we know the final result.
However, there is no justification for converting a covariant action into
non-covariant form before starting the Hamiltonian procedure. In particular,
it is not apparent at all why one would do this in the Hamiltonian formulation
of linearized GR/spin-2. Here we can only guess answers to the above
questions. This non-covariant modification of the action is likely related to
Dirac's work on the Hamiltonian formulation of full GR \cite{Dirac} which came
before the Hamiltonian formulation of the weak approximation appeared in the
literature. Dirac explained that the reason for this modification is to
simplify the primary constraints. However, he warned that these
simplifications \textquotedblleft force one to abandon the four-dimensional
symmetry\textquotedblright. This change in the Lagrangian occurs in equation
(15) of \cite{Dirac}:%

\begin{equation}
L_{\Gamma\Gamma}\rightarrow L_{\Gamma\Gamma}^{\ast}=L_{\Gamma\Gamma}-\left[
\left(  \sqrt{-g}g^{00}\right)  _{,k}\frac{g^{k0}}{g^{00}}\right]
_{,0}+\left[  \left(  \sqrt{-g}g^{00}\right)  _{,0}\frac{g^{k0}}{g^{00}%
}\right]  _{,k}. \label{eq.9}%
\end{equation}
Note that here and everywhere we have chosen the opposite sign to Dirac
because we use the definition of $L$ in \cite{Landau, Carmeli}. (Here Latin
letters stand for space indices ($k=1,2,3,...$) and \textquotedblleft%
$0$\textquotedblright\ for the time index).

The last two non-covariant terms in (\ref{eq.9}) do not affect the equations
of motion and simplify the primary constraints (see the next Section) because
they eliminate terms linear in the velocities $g_{0\mu,0}$. In the linearized
case (which can be obtained by linearization of (\ref{eq.9}) or by
linearization of the general result of Dirac's (see equation (17) of
\cite{Dirac}) these two terms become%

\[
L_{\Gamma\Gamma}^{\ast}-L_{\Gamma\Gamma}=\frac{1}{2}\eta^{00}\left[
h_{00,0}h_{m0,k}\eta^{mk}\eta^{00}-h_{pq,0}h_{m0,k}\eta^{pq}\eta^{mk}%
-h_{m0,0}h_{00,k}\eta^{mk}\eta^{00}+h_{m0,0}h_{pq,k}\eta^{pq}\eta^{mk}\right]
\]

\begin{equation}
=\frac{1}{2}\left[  h_{00,0}h_{0,k}^{k}+h_{p,0}^{p}h_{0,k}^{k}-h_{0,0}%
^{k}h_{00,k}-h_{0,0}^{k}h_{p,k}^{p}\right]  . \label{eq.10}%
\end{equation}

Equation (\ref{eq.10}) in combination with $L_{\Gamma\Gamma}$ leads to a
cancellation of some terms that are linear in the velocities, and results in
Dirac's Lagrangian as used in \cite{Baaklini, Ferrari}%

\begin{align}
L_{Dirac\left(  lin\right)  }  &  =\frac{1}{4}h_{~,0}^{pq}h_{pq,0}-\frac{1}%
{4}h_{p,0}^{p}h_{q,0}^{q}+h_{p,0}^{p}h_{0,k}^{k}-h_{~,0}^{pk}h_{0p,k}+\frac
{1}{2}h_{00,k}\left(  h_{~,p}^{kp}-\eta^{kp}h_{q,p}^{q}\right)  +\label{LD}\\
&  \frac{1}{2}h_{0,k}^{p}h_{0p}^{~,k}-\frac{1}{2}h_{0,k}^{p}h_{0,p}^{k}%
+\frac{1}{4}h_{p,k}^{p}h_{q}^{q,k}-\frac{1}{4}h_{~,k}^{pq}h_{pq}^{~,k}%
+\frac{1}{2}h_{~,k}^{pq}h_{p,q}^{k}-\frac{1}{2}h_{~,k}^{pk}h_{q,p}%
^{q}.\nonumber
\end{align}

Later we will demonstrate explicitly that it is \textit{not} necessary to make
this non-covariant modification in the Hamiltonian formulation of linearized
gravity. Then the relevant discussion about the necessity of this modification
for full GR will be presented in Discussion.

\section{Hamiltonian}

\subsection{Non-covariant case}

The method used here was developed by Dirac \cite{Diracbook} and below we
briefly summarize the main steps of his procedure. Introducing the momenta
conjugate to $h_{\mu\nu}$:%

\begin{equation}
p^{\mu\nu}=\frac{\delta L_{Dirac(lin)}}{\delta h_{\mu\nu,0}} \label{eq.12}%
\end{equation}
gives the primary constraints%

\begin{equation}
\phi^{0\mu}=p^{0\mu}\approx0, \label{eq.13}%
\end{equation}
and an equation for $p^{kn}$
\begin{equation}
p^{kn}=\frac{1}{2}\eta^{kn}h_{p}^{p,0}-\frac{1}{2}h^{kn,0}+\frac{1}{2}%
h^{k0,n}+\frac{1}{2}h^{n0,k}-\eta^{kn}h_{~,p}^{p0}~. \label{eq.14}%
\end{equation}
Solving (\ref{eq.14}) for the velocities gives%

\begin{equation}
h_{kn,0}=2p_{kn}-\frac{2}{d-2}\eta_{kn}p_{m}^{m}+h_{0k,n}+h_{0n,k}.
\label{eq.15}%
\end{equation}

The factor of $(d-2)$ (where $d$ is the dimension of spacetime) appearing in
the denominator reflects the fact that (\ref{eq.14}) cannot be solved for a
velocity when $d=2$. The treatment of two-dimensional linearized gravity is
given separately in the following Subsection.

Substituting velocities from (\ref{eq.15}) into the Lagrangian (\ref{LD}), we
obtain the canonical Hamiltonian%

\[
H_{c(Dirac)}=p^{kn}\dot{h}_{kn}-L_{lin}=p_{n}^{k}p_{k}^{n}-\frac{1}{d-2}%
p_{k}^{k}p_{n}^{n}+2p^{nk}h_{k0,n}-\frac{1}{2}h_{00,k}\left(  h_{,p~}%
^{kp}-\eta^{kp}h_{q,p}^{q}\right)
\]

\begin{equation}
-\frac{1}{4}h_{p,k}^{p}h_{q}^{q,k}+\frac{1}{4}h_{~,k}^{pq}h_{pq}^{~,k}%
-\frac{1}{2}h_{,k~}^{pq}h_{p,q}^{k}+\frac{1}{2}h_{~,k}^{pk}h_{q,p}^{q}
\label{eq.16}%
\end{equation}
and the total Hamiltonian%

\begin{equation}
H_{T(Dirac)}=H_{c(Dirac)}+\dot{h}_{0\mu}\phi^{0\mu}. \label{eq.17}%
\end{equation}

Direct calculation of the Hamiltonian (\ref{eq.16}) results in two additional
terms, $h_{0,k}^{p}h_{0,p}^{k}$ and $-h_{0,p}^{p}h_{0,k}^{k}$, but since these
two terms can be expressed as a total derivative, we do not include them in
the Hamiltonian.

With the fundamental Poisson brackets (PB) defined as\footnote{We will omit
$\delta_{d-1}\left(  x-x^{\prime}\right)  $ in further calculations.}%

\begin{equation}
\left\{  h_{\alpha\beta}\left(  x\right)  ,p^{\mu\nu}\left(  x^{\prime
}\right)  \right\}  =\frac{1}{2}\left(  \delta_{\alpha}^{\mu}\delta_{\beta
}^{\nu}+\delta_{\alpha}^{\nu}\delta_{\beta}^{\mu}\right)  \delta_{d-1}\left(
x-x^{\prime}\right)  \equiv\Delta_{\alpha\beta}^{\mu\nu}\delta_{d-1}\left(
x-x^{\prime}\right)  ~, \label{eq.17a}%
\end{equation}
\qquad the conservation in time of primary (\ref{eq.13}) constraints leads to
the secondary constraints%

\begin{equation}
\dot{\phi}^{00}=\left\{  \phi^{00},H_{T}\right\}  =\frac{1}{2}\left(
h_{k,n}^{k,n}-h_{n,k}^{k,n}\right)  \equiv\chi^{00}, \label{eq.18}%
\end{equation}

\begin{equation}
\dot{\phi}^{0n}=\left\{  \phi^{0n},H_{T}\right\}  =p_{~,k~}^{nk}\equiv
\chi^{0n}. \label{eq.19}%
\end{equation}

All constraints have vanishing PB among themselves, so all of them are first
class (FC) at this stage of the Dirac procedure. To verify the closure of the
Dirac procedure we have to consider the time development of the secondary
constraints to check whether they produce any new constraints. We find%
\begin{equation}
\dot{\chi}^{00}=\left\{  \chi^{00},H_{T}\right\}  =-\chi_{~,k}^{0k},\quad
\dot{\chi}^{0n}=\left\{  \chi^{0n},H_{T}\right\}  =0, \label{eq.20a}%
\end{equation}
so, no new constraints appear and the Dirac procedure closes with $2d$ FC
constraints ($d>2$). Counting the degrees of freedom gives, for example, in
the $d=4$ case: $10~(variables~h_{\mu\nu})-8~(FC~constraints)=2$, as is
expected for the massless spin-2 system.

Now, from our knowledge of the FC constraints, we can find the gauge
transformations by using the procedure of Castellani \cite{Castellani}. The
gauge generators are of the form:%

\begin{align}
G_{(0)}  &  =-\chi^{00}+\int\left(  \alpha\phi^{00}+\alpha_{n}\phi
^{0n}\right)  d^{d-1}x,\\
G_{(0)}^{n}  &  =-\chi^{0n}+\int\left(  \beta^{n}\phi^{00}+\beta\phi
^{0n}\right)  d^{d-1}x. \nonumber\label{eq.20}%
\end{align}

The coefficients $\alpha,\alpha_{n}$ and $\beta,\beta^{n}$ can be found from
the conditions \cite{Castellani}:%
\[
\dot{G}_{(0)}=\left\{  G_{(0)},H\right\}  =\chi_{~,n}^{0n}+\alpha\chi
^{00}+\alpha_{n}\chi^{0n}=0\quad\Rightarrow\quad\alpha=0,\quad\alpha
_{n}=-\partial_{n},\newline%
\]%
\begin{equation}
\dot{G}_{(0)}^{n}=\left\{  G_{(0)}^{n},H\right\}  =0+\beta^{n}\chi^{00}%
+\beta\chi^{0n}=0\quad\Rightarrow\quad\beta^{n}=\beta=0. \label{eq.21}%
\end{equation}

The total gauge generator is%

\begin{align}
G\left(  \varepsilon_{\gamma},\dot{\varepsilon}_{\gamma}\right)   &  =\int
d^{d-1}x\left(  \varepsilon_{0}G_{(0)}+\dot{\varepsilon}_{0}\phi
^{00}+\varepsilon_{n}G_{(0)}^{n}+\dot{\varepsilon}_{n}\phi^{0n}\right)
=\nonumber\\
&  \int d^{d-1}x\left(  -\varepsilon_{0}\chi^{00}-\varepsilon_{0}\phi
_{~,n}^{0n}+\dot{\varepsilon}_{0}\phi^{00}-\varepsilon_{n}\chi^{0n}%
+\dot{\varepsilon}_{n}\phi^{0n}\right)  , \label{eq.22}%
\end{align}
where $\varepsilon_{\mu}$ is the gauge parameter. It is easy to show that the
PB of two generators is zero since PBs among all FC constraints are zero. Now
we can find the gauge transformations of the canonical variables $h_{\mu\nu}$
and $p^{\mu\nu}$:%

\begin{equation}
\delta h_{\mu\nu}=\left\{  G\left(  \varepsilon_{\gamma},\dot{\varepsilon
}_{\gamma}\right)  ,h_{\mu\nu}\right\}  =-\frac{1}{2}\left(  \varepsilon
_{\mu,\nu}+\varepsilon_{\nu,\mu}\right)  , \label{eq.23}%
\end{equation}

\begin{equation}
\delta p^{00}=\delta p^{0k}=0,\quad\delta p^{kn}=\frac{1}{2}\left(  \eta
^{kn}\varepsilon_{0,p}^{~,p}-\varepsilon_{0}^{~,kn}\right)  , \label{eq.24}%
\end{equation}
which give $\delta\chi^{0n}=\delta p_{~,k}^{kn}=0$ and $\delta\chi^{00}=0$. It
is straightforward to verify that these gauge transformations keep the
Lagrangian invariant up to a total derivative. Note that the final expression
(\ref{eq.23}) \ has four-dimensional (or $d-$dimensional, in general) symmetry
and Dirac's statement about \textquotedblleft abandoning four-dimensional
symmetry\textquotedblright\ is restricted to only his modification of the
Lagrangian which initially possesses this four-dimensional symmetry.

When $d=2$, Dirac's modified Lagrangian $L_{Dirac\left(  lin\right)  }$
vanishes, so we can say that the theory is meaningless in two dimensions or,
using the words of Jackiw, \textquotedblleft it cannot even be
formulated\textquotedblright\ \cite{Jackiw-1}. However, from the fact that the
modified Lagrangian is zero, it does not follow that the EH Lagrangian is
meaningless in two dimensions. It follows that the non-covariant modification
of (\ref{eq.9}) should not be performed because it eliminates the essential
contributions to the original Lagrangian. In contrast, if we do not modify the
Lagrangian, then the Hamiltonian analysis gives consistent results in both the
linearized (see the next Subsection) and the non-linearized two-dimensional
cases \cite{OnHam, Two-dim}.

\subsection{Covariant case}

We redo the Hamiltonian formulation for the massless spin-2 field, but now use
the covariant form (\ref{eq.1}) subject to the conditions (\ref{eq.3}). To be
able to compare our results with the results of others and to see what effects
(if any) the presence of a free parameter produces, we perform our
calculations using the most general covariant Lagrangian:%

\begin{equation}
L=\frac{1}{4}h_{\alpha\beta,\gamma}h_{\mu\nu,\rho}\left[  \eta^{\alpha\beta
}\eta^{\mu\nu}\eta^{\gamma\rho}-\eta^{\alpha\mu}\eta^{\beta\nu}\eta
^{\gamma\rho}+c_{3}\eta^{\alpha\gamma}\eta^{\beta\nu}\eta^{\mu\rho}+c_{4}%
\eta^{\alpha\rho}\eta^{\beta\nu}\eta^{\gamma\mu}-2\eta^{\alpha\beta}%
\eta^{\gamma\mu}\eta^{\nu\rho}\right]  \label{star}%
\end{equation}
keeping the relation $c_{3}+c_{4}=2$.

Introducing momenta conjugate to $h_{\mu\nu}$:%

\begin{equation}
\pi^{\mu\nu}=\frac{\delta L}{\delta h_{\mu\nu,0}} \label{eq.25}%
\end{equation}
gives the primary constraints%

\begin{equation}
\phi^{00}=\pi^{00}-\frac{1}{2}\left(  1-c_{3}\right)  h_{~,k}^{0k},
\label{eq.26}%
\end{equation}

\begin{equation}
\phi^{0n}=\pi^{0n}+\frac{c_{3}}{4}h_{~,k}^{nk}-\frac{1}{4}h_{k}^{k,n}%
+\frac{c_{3}-1}{4}h^{00,n}. \label{eq.27}%
\end{equation}

In the covariant case the maximum simplification of the primary constraints
follows from setting $c_{3}=1,$ so that one constraint becomes $\pi
^{00}\approx0$. However, we cannot do this in the full GR Lagrangian without
destroying the symmetry of $d-$dimensional spacetime.

From $\pi^{kn}=\frac{\delta L}{\delta h_{kn,0}}$ one obtains
\begin{equation}
\pi^{kn}=\frac{1}{4}\left(  2\eta^{kn}h_{p}^{p,0}-2h^{kn,0}+c_{4}\left(
h^{k0,n}+h^{n0,k}\right)  -2\eta^{kn}h_{~,p}^{p0}\right)  . \label{eq.28}%
\end{equation}
Solving (\ref{eq.28}) for the velocities gives the following equation%

\begin{equation}
h_{kn,0}=2\pi_{kn}-\frac{2}{d-2}\eta_{kn}\pi_{p}^{p}+\frac{1}{d-2}\left(
1-c_{4}\right)  \eta_{kn}h_{0,p}^{p}+\frac{c_{4}}{2}\left(  h_{0k,n}%
+h_{0n,k}\right)  . \label{eq.29}%
\end{equation}
Once again, the factor $(d-2)$ appears in the denominator of some of the terms
in the right-hand side of (\ref{eq.29}) for the same reason as in the previous Subsection.

Substituting the velocities from (\ref{eq.29}) into the Lagrangian
(\ref{star}) one obtains the canonical Hamiltonian%

\[
H_{c}=\pi^{kn}\dot{h}_{kn}-L=\pi_{n}^{k}\pi_{k}^{n}-\frac{1}{d-2}\pi_{k}%
^{k}\pi_{n}^{n}-c_{4}\pi_{k}^{n}h_{~,n}^{0k}-\frac{1-c_{4}}{d-2}\pi_{k}%
^{k}h_{~,n}^{0n}+
\]

\begin{equation}
\left[  \frac{1-c_{4}}{4}-\frac{\left(  1-c_{4}\right)  ^{2}}{4\left(
d-2\right)  }\right]  h_{~,k}^{0k}h_{~,n}^{0n}+\left(  \frac{\left(
c_{4}\right)  ^{2}}{8}+\frac{c_{4}}{4}\right)  h_{~,n}^{0k}h_{~,k}%
^{0n}+\left(  \frac{\left(  c_{4}\right)  ^{2}}{8}-\frac{1}{2}\right)
h_{~,n}^{0k}h_{k}^{0,n}+ \label{eq.30}%
\end{equation}

\[
\frac{1}{2}h_{00}^{~,n}\left(  h_{k,n}^{k}-h_{n,k}^{k}\right)  -\frac{1}%
{4}h_{n,k}^{n}h_{p}^{p,k}+\frac{1}{4}h_{p,k}^{n}h_{n}^{p,k}-\frac{c_{3}}%
{4}h_{pn}^{~,n}h_{~,k}^{kp}-\frac{c_{4}}{4}h_{n}^{p,k}h_{pk}^{~,n}+\frac{1}%
{2}h_{p,k}^{p}h_{~,n}^{nk}%
\]
and the total Hamiltonian%

\begin{equation}
H_{T}=H_{c}+\dot{h}_{0\mu}\phi^{0\mu}. \label{eq.30a}%
\end{equation}

Setting $c_{3}=0$ and $c_{4}=2$ in (\ref{eq.30}) leaves us with a couple more
terms than in the non-covariant case of (\ref{eq.16}), more specifically:%
\[
H_{c(covariant,~c_{3}=0)}-H_{c(Dirac)}=\frac{1}{d-2}\pi_{k}^{k}h_{\ ,n}%
^{0n}-\frac{1}{4}\left(  \frac{1}{d-2}-3\right)  h_{\ ,k}^{0k}h_{\ ,n}^{0n}%
\]
where a total derivative has again been omitted.

With the fundamental PB defined as%

\[
\left\{  h_{\alpha\beta},\pi^{\mu\nu}\right\}  =\Delta_{\alpha\beta}^{\mu\nu
}~,
\]
the conservation of primary constraints in time gives the secondary constraints%

\begin{equation}
\dot{\phi}^{00}=\left\{  \phi^{00},H_{T}\right\}  =\frac{1}{2}\left(
h_{k,n}^{k,n}-h_{n,k}^{k,n}\right)  \equiv\chi^{00}, \label{eq.31}%
\end{equation}

\begin{equation}
\dot{\phi}^{0n}=\left\{  \phi^{0n},H_{T}\right\}  =\pi_{~,k}^{nk}+\frac{c_{4}%
}{4}h_{0,k}^{k,n}-\frac{c_{3}}{4}h_{0,k}^{n,k}\equiv\chi^{0n}. \label{eq.32}%
\end{equation}

Note that in the covariant case, even for linearized gravity, $\chi^{0n}$
depends on the spatial derivatives of $h_{0k}$.

All constraints have zero PBs among themselves, so all of them are first class
(FC) at this stage of the Dirac procedure. It is easy to verify that no new
constraints appear:%

\begin{equation}
\dot{\chi}^{00}=\left\{  \chi^{00},H_{T}\right\}  =-\chi_{\ ,k}^{0k},\quad
\dot{\chi}^{0n}=\left\{  \chi^{0n},H_{T}\right\}  =0. \label{eq.32a}%
\end{equation}

Despite the difference in the expressions of the primary and secondary
constraints, the constraint structure and the number of degrees of freedom are
the same as in the non-covariant case (\ref{eq.20a}).

From the FC constraint structure, we can once again find the gauge
transformations using the Castellani procedure, described in the previous Subsection.

The total gauge generator can be calculated as in (\ref{eq.22}), except that
the constraints are now given by (\ref{eq.26}), (\ref{eq.27}), (\ref{eq.31}),
and (\ref{eq.32}). Nevertheless, it produces the same gauge transformations
for $h_{\mu\nu}$

\[
\delta h_{\mu\nu}=\left\{  G\left(  \varepsilon_{\gamma},\dot{\varepsilon
}_{\gamma}\right)  ,h_{\mu\nu}\right\}  =-\frac{1}{2}\left(  \varepsilon
_{\mu,\nu}+\varepsilon_{\nu,\mu}\right)
\]
while for $\pi^{\mu\nu}$ it gives%

\begin{align}
\delta\pi^{00}  &  =\frac{c_{3}-c_{4}}{8}\left(  \varepsilon_{0,n}^{~,n\quad
}+\varepsilon_{n,0}^{~,n}\right)  ,\quad\delta\pi^{0k}=\frac{c_{3}-c_{4}}%
{8}\varepsilon_{0}^{~,0k}+\frac{c_{4}}{8}\varepsilon_{~,n}^{n,k}-\frac{c_{3}%
}{8}\varepsilon_{~~,n}^{k,n}~,\\
\delta\pi^{kn}  &  =\frac{1}{4}\eta^{kn}\varepsilon_{0,p}^{~,p}-\frac{1}%
{4}\eta^{kn}\varepsilon_{p,0}^{~,p}-\frac{c_{4}}{4}\varepsilon_{0}%
^{~,kn}+\frac{c_{3}}{8}\left(  \varepsilon_{~~,0}^{k,n}+\varepsilon
_{~~,0}^{n,k}\right)  . \nonumber\label{eq.33}%
\end{align}

It is straightforward to verify that these gauge transformations, as in the
non-covariant case, leave all constraints unchanged and the Lagrangian
invariant up to a total derivative:%

\[
\delta L
%\frac{1}{4}\left[  c_{3}\partial_{\alpha}\left(  \varepsilon
%^{\gamma,\alpha}h_{\mu\gamma}^{\quad,\mu}\right)  -c_{3}\partial^{\mu}\left(
%\varepsilon^{\gamma,\alpha}h_{\mu\gamma,\alpha}\right)  +c_{4}\partial^{\mu
%}\left(  \varepsilon^{\gamma,\alpha}h_{\alpha\mu,\gamma}\right)
%-c_{4}\partial_{\gamma}\left(  \varepsilon^{\gamma,\alpha}h_{\alpha\mu}%
%^{\quad,\mu}\right)  \right]
%\]
=\frac{1}{4}\left[  \partial_{\alpha}\left(  \left(  c_{3}\varepsilon
^{\gamma,\alpha}-c_{4}\varepsilon^{\alpha,\gamma}\right)  h_{\mu\gamma
}^{~~,\mu}\right)  -\partial^{\mu}\left(  \left(  c_{3}\varepsilon
^{\gamma,\alpha}-c_{4}\varepsilon^{\alpha,\gamma}\right)  h_{\gamma\mu,\alpha
}\right)  \right]  .
\]

When $d=2$, the original Lagrangian of (\ref{star}) is drastically simplified
(note, if $c_{4}=2$, this Lagrangian can be obtained from the two-dimensional
Lagrangian given in \cite{OnHam} by applying the substitution of (\ref{eq.8a})
to $L_{\Gamma\Gamma}$):%

\begin{equation}
L_{2d}=\frac{c_{3}-c_{4}}{4}\left(  h_{11,1}h_{01,0}-h_{01,1}h_{11,0}%
+h_{00,1}h_{01,0}-h_{01,1}h_{00,0}\right)  . \label{eq.34}%
\end{equation}

This expression is linear in the velocities, so no velocity can be eliminated
using the equations of motion. The momenta conjugate to $h_{00}$, $h_{11}$,
and $h_{01}$ give three primary constraints:%

\begin{align}
\phi^{00}  &  =\pi^{00}+\frac{c_{3}-c_{4}}{4}h_{01,1}~,\quad\phi^{11}=\pi
^{11}+\frac{c_{3}-c_{4}}{4}h_{01,1}~,\label{eq.41}\\
\phi^{01}  &  =\pi^{01}-\frac{c_{3}-c_{4}}{4}\left(  h_{11,1}+h_{00,1}\right)
. \nonumber\label{eq.35}%
\end{align}

The PBs among these constraints are all zero, and their time development does
not produce any new constraints. Therefore the Dirac procedure closes with
three FC primary constraints, so that there are zero degrees of freedom when
$d=2$. The canonical Hamiltonian is zero and the total Hamiltonian is then
just a linear combination of primary constraints:%

\begin{equation}
H_{T}=\dot{h}_{\mu\nu}\phi^{\mu\nu}. \label{eq.41a}%
\end{equation}

The gauge generator is simplified as well%

\begin{equation}
G=\int dx\left(  \varepsilon_{00}\phi^{00}+2\varepsilon_{01}\phi
^{01}+\varepsilon_{11}\phi^{11}\right)  ~, \label{eq.36}%
\end{equation}
which results in the following gauge transformations of the canonical variables%

\begin{equation}
\delta h_{\mu\nu}=-\varepsilon_{\mu\nu}~, \label{eq.37}%
\end{equation}

\begin{equation}
\delta\pi^{00}=\delta\pi^{11}=-\frac{c_{4}-c_{3}}{2}\varepsilon_{01,1}%
~,\quad\delta\pi^{01}=\frac{c_{4}-c_{3}}{4}\left(  \varepsilon_{00,1}%
+\varepsilon_{11,1}\right)  . \label{eq.38}%
\end{equation}

One should note that the gauge transformations of $h_{\mu\nu}$ for $d=2$ are
exactly the same as were obtained for the full (not linearized)
\textquotedblleft gamma-gamma\textquotedblright\ part of the EH Lagrangian in
\cite{OnHam} and for the full EH Lagrangian using the Ostrogradsky procedure
in \cite{Two-dim}. These gauge transformations are consistent with the
triviality of the Einstein equations in two dimensions. (In \cite{OnHam} it is
pointed out that in gauge in which $g_{01}\neq0$, the EH action when $d=2$
leads to the meaningful Hamiltonian formulation).

The gauge transformations for $h_{\mu\nu}$ in (\ref{eq.37}) leave the
Lagrangian invariant up to a total derivative, which can be cast in a
covariant form%

\[
\delta L_{2d}=\frac{c_{3}-c_{4}}{4}\epsilon^{\alpha\beta}\epsilon^{\gamma\rho
}\eta^{\mu\nu}\partial_{\beta}\left(  h_{\gamma\mu}\varepsilon_{\rho\nu
,\alpha}\right)
\]
where $\epsilon^{\alpha\beta}=-\epsilon^{\beta\alpha}$.

It is easy to check that $\delta\phi^{\mu\nu}=0$ and $\delta H_{T}=0$ under
these gauge transformations.

We note that if $c_{3}=c_{4}=1$, then the Lagrangian (\ref{eq.34}) is zero,
and therefore in this case there is no Hamiltonian formulation. These values
of $c_{3}$ and $c_{4}$ are only possible in linearized GR. The full GR in
covariant form requires $c_{3}=0$ and $c_{4}=2$ as was mentioned above.

\section{Equivalence}

The equivalence of different formulations of linearized GR/spin-2 considered
in previous sections can be discussed for the Lagrangians and the
corresponding Hamiltonians. At the Lagrangian level this equivalence is quite
obvious because the parameters of the covariant spin-2 Lagrangian (\ref{eq.1})
were found from the condition of invariance given in (\ref{eq.2}) and two
terms proportional to $c_{3}$ and $c_{4}$ are equivalent up to total
divergences of (\ref{eq.4}) and do not affect the equations of motion. Dirac's
non-covariant modifications make (\ref{eq.9}) differ from (\ref{eq.8}) by
non-covariant integrations, however, the Dirac Lagrangian also leads to the
same equations of motion. We thus can conclude that both Lagrangians are equivalent.

The demonstration of equivalence for the Hamiltonian formulations in the
covariant and modified by Dirac cases is a little bit more involved and worth
discussing in some detail. We have found, that despite having different
primary and secondary constraints and different expressions for the canonical
Hamiltonians, the PBs among the constraints and constraints with the
Hamiltonian have the same structure for both formulations (see (\ref{eq.20a}),
(\ref{eq.32a})). The Castellani procedure also leads to the same gauge transformations.

From ordinary classical mechanics \cite{Lan} it is known that performing
canonical transformations from one set of variables to another leads to
equivalent Hamiltonian formulations. Let us try to find such transformations
between the covariant $(h,\pi)$ variables and the $(h,p)$ of Dirac's
formulation. The change of the momenta $p^{0\mu}$ is quite obvious%

\begin{equation}
p^{00}=\pi^{00}-\frac{1-c_{3}}{2}h_{~~,k}^{0k}, \label{eq.42}%
\end{equation}

\begin{equation}
p^{0n}=\pi^{0n}+\frac{c_{3}}{4}h_{~,k}^{nk}-\frac{1}{4}h_{k}^{k,n}+\frac
{c_{3}-1}{4}h^{00,n}. \label{eq.43}%
\end{equation}
For the space-space components of the momenta one can try to obtain such
relations by comparing expressions for the corresponding velocities
((\ref{eq.15}) and (\ref{eq.29})) in the two formulations which gives%

\begin{equation}
p^{nk}=\pi^{nk}+\frac{c_{3}}{4}\left(  h^{k0,n}+h^{0n,k}\right)  -\frac{1}%
{2}\eta^{kn}h_{~~,p}^{p0}. \label{eq.44}%
\end{equation}
Generalized coordinates are treated in the same way in both formulations, so
we have the following transformations%

\begin{equation}
p^{\alpha\beta}=p^{\alpha\beta}\left(  \pi^{\mu\nu},h_{\mu\nu}\right)
,\left.  {}\right.  \left.  {}\right.  h_{\alpha\beta}=h_{\alpha\beta}.
\label{eq.45}%
\end{equation}
To check whether the change of variables is canonical, one has to verify that
the following PBs are satisfied \cite{Lan}%

\begin{equation}
\left\{  h_{\mu\nu},p^{\alpha\beta}\right\}  _{p,h}=\left\{  h_{\mu\nu
},p^{\alpha\beta}\left(  \pi^{\mu\nu},h_{\mu\nu}\right)  \right\}  _{\pi
,h}=\Delta_{\mu\nu}^{\alpha\beta}~, \label{eq.46}%
\end{equation}

\begin{equation}
\left\{  p^{\alpha\beta},p^{\mu\nu}\right\}  _{p,h}=\left\{  p^{\alpha\beta
},p^{\mu\nu}\right\}  _{\pi,h}=0 \label{eq.47}%
\end{equation}
and%

\begin{equation}
\left\{  h_{\alpha\beta},h_{\mu\nu}\right\}  _{p,h}=\left\{  h_{\alpha\beta
},h_{\mu\nu}\right\}  _{\pi,h}=0. \label{eq.48}%
\end{equation}

Equation (\ref{eq.48}) is obviously satisfied by (\ref{eq.45}), and equations
(\ref{eq.46}) and (\ref{eq.47}) can be shown to be satisfied after a short
calculation. For Hamiltonian formulations of non-singular Lagrangians
demonstrating (\ref{eq.46})-(\ref{eq.48}) would be enough to prove that the
two Hamiltonians are equivalent. For singular models, this is only a necessary
condition. To show equivalence for the case of two Hamiltonian formulations of
a gauge invariant Lagrangian, we must also demonstrate that the whole algebra
of constraints is preserved because the whole algebra is needed to find the
generator of gauge transformations. Let us check this requirement and
substitute the inverse of (\ref{eq.45}), $\pi^{\alpha\beta}=\pi^{\alpha\beta
}\left(  p^{\mu\nu},h_{\mu\nu}\right)  $, into the total Hamiltonian in the
covariant formulation of (\ref{eq.30a})%

\[
H_{T}\left(  \pi^{\alpha\beta},h_{\alpha\beta};c_{3},c_{4}\right)  =\dot
{h}_{0\mu}\phi^{0\mu}+H_{c}\left(  \pi^{km},h_{\alpha\beta};c_{3}%
,c_{4}\right)  =
\]

\[
\dot{h}_{0\mu}p^{0\mu}+p_{n}^{k}p_{k}^{n}-\frac{1}{d-2}p_{k}^{k}p_{n}%
^{n}+2p^{nk}h_{k0,n}-\frac{1}{2}h_{00,k}\left(  h_{~,p~}^{kp}-\eta^{kp}%
h_{q,p}^{q}\right)
\]

\begin{equation}
-\frac{1}{4}h_{p,k}^{p}h_{q}^{q,k}+\frac{1}{4}h_{~~,k}^{pq}h_{pq}%
^{~~~,k}-\frac{c_{3}}{4}h_{pq}^{~~~,q}h_{~~,k}^{kp}-\frac{c_{4}}{4}%
h_{~~,k~}^{pq}h_{p,q}^{k}+\frac{1}{2}h_{~~,k}^{pk}h_{q,p}^{q}~. \label{eq.49}%
\end{equation}

With $c_{3}=0$ ($c_{4}=2$) we recover the Hamiltonian of Dirac (\ref{eq.17})
that was obtained by the non-covariant modification of the Lagrangian for
linearized GR. In fact, according to (\ref{eq.49}), similar modifications in
the linearized GR/spin-2 case can be made for all possible values of
parameters satisfying $c_{3}+c_{4}=2$. Hence, the Hamiltonian formulations
that generate the same gauge transformations are related by canonical
transformations that also preserve the whole algebra of constraints, although
the expressions of the constraints themselves do change.

We will now briefly comment on similar relations when $d=2$. In this case, we
consider the following change of variables, using (\ref{eq.41}),%

\begin{equation}
p^{00}=\pi^{00}+\frac{c_{3}-c_{4}}{4}h_{01,1}~,\quad p^{11}=\pi^{11}%
+\frac{c_{3}-c_{4}}{4}h_{01,1}~,\quad p^{01}=\pi^{01}-\frac{c_{3}-c_{4}}%
{4}\left(  h_{11,1}+h_{00,1}\right)  . \label{eq.50}%
\end{equation}
which satisfy the conditions (\ref{eq.46})-(\ref{eq.48}). Substitution of
(\ref{eq.50}) into (\ref{eq.41a}) gives%

\begin{equation}
H_{T}=\dot{h}_{00}p^{00}+2\dot{h}_{01}p^{01}+\dot{h}_{11}p^{11}. \label{eq.51}%
\end{equation}

We see that using the Castellani procedure, one finds the same gauge
transformations as in (\ref{eq.37}). Keeping the two-dimensional Lagrangian
(\ref{eq.34}) as it is and performing canonical transformations at the
Hamiltonian level gives another consistent Hamiltonian formulation of
linearized GR when $d=2$ (with simple constraints). This is different from the
case where Dirac's modifications were performed at the Lagrangian level and
led to the disappearance of the Lagrangian in two dimensions. One may say that
two-dimensional GR makes no sense and that discussing its Hamiltonian
formulation without modifications of the Lagrangian is meaningless
\cite{Jackiw-1}. However, exactly the opposite is correct and without
non-covariant modifications of the Lagrangian we have meaningful Hamiltonian
formulation of two-dimensional GR.

In the covariant case we have one set of parameter values ($c_{3}=c_{4}=1$)
that also leads to complete disappearance of the Lagrangian, but we can pick
these values \emph{only} in the linearized EH action. For full GR only one
value is permissible, $c_{4}=2$, if one is to keep the invariance of the
action under general coordinate transformations.

A possible objection against our result for the Hamiltonian formulation of
two-dimensional GR is that it does not produce the \textquotedblleft
expected\textquotedblright\ gauge invariance, diffeomorphism. However, when
$d=2$ it is not possible to obtain diffeomorphism as a gauge symmetry using
the Dirac procedure. This conclusion can be drawn from a simple consideration:
to construct the diffeomorphism transformation of $g_{\alpha\beta}$, using,
for example, the Castellani procedure, one needs $d$ primary and $d$ secondary
FC constraints. In two dimensions this gives four FC constraints. However,
when $d=2$, a metric tensor has only three independent components and counting
the degrees of freedom leads to minus one, meaning that the system is
overconstrained, and non-physical. We assert that the effects of non-covariant
modifications of the EH Lagrangian as well as canonical transformations among
different sets of variables used in the Hamiltonian formulations of GR should
be investigated.

\section{Discussion}

To draw conclusions about the relevance of linearized GR/spin-2 to the full
theory, we compare the Hamiltonian formulation given in this paper with the
Hamiltonian formulations of full GR for both cases: covariant and with Dirac's
modifications. In \cite{PLA} we presented the complete Hamiltonian analysis,
including the restoration of gauge invariance, of the EH action for full GR
without any modifications or change of variables. (This analysis was started
in 1952 by Pirani, Schild and Skinner (PSS) in \cite{Pirani} but had never
been completed before.) In \cite{myths} similar analysis was carried out for
the Dirac Lagrangian of \cite{Dirac}. In both formulations (PSS and Dirac)
despite having a different form for the constraints and Hamiltonians, the
algebra of constraints remains the same and the generators of the gauge
transformations built from the first class constraints produce the
diffeomorphism transformation of the metric tensor. Moreover, in \cite{canon}
we showed that these two formulations (PSS and Dirac) are related to each
other by a canonical transformation, in contrast to the ADM formulation
\cite{ADM} which is not related to either by any canonical transformation (for
the proof see Sec. 4 of \cite{myths}).

We can point out some similarities in the Hamiltonian analysis for the full
and linearized EH action with the metric tensor as a fundamental variable: the
number of constraints, their algebra, and the gauge transformations of the
metric tensor are in exact correspondence. It is possible to show that the
linearization of the constraints in the Dirac formulation \cite{myths}
produces exactly the expressions (\ref{eq.26}), (\ref{eq.27}), (\ref{eq.31}),
(\ref{eq.32}), and linearization of the covariant case \cite{PLA} produces
exactly (\ref{eq.13}), (\ref{eq.18}), (\ref{eq.19}). Time development of the
secondary first class constraints gives (\ref{eq.32a}) and (\ref{eq.20a}) upon
linearization in both cases. Finally, linearization of the gauge
transformation of the metric tensor leads to (\ref{eq.23})\footnote{In making
the comparison, one has to take into account that Dirac in \cite{Dirac} used
the convention $\left\{  p^{\mu\nu}\left(  x\right)  ,h_{\alpha\beta}\left(
x^{\prime}\right)  \right\}  =\frac{1}{2}\left(  \delta_{\alpha}^{\mu}%
\delta_{\beta}^{\nu}+\delta_{\alpha}^{\nu}\delta_{\beta}^{\mu}\right)
\delta_{3}\left(  x-x^{\prime}\right)  $ which differs from (\ref{eq.17a}) and
will give an opposite sign, for example in constraints, if the constraints of
the Dirac formulation of \cite{Dirac} are linearized (see \cite{myths} for
details). Note that in \cite{canon} we employ the convention (\ref{eq.17a}) in
both formulations (Dirac's and PSS).}.

We would like to note that we were not able to compare the full ADM
formulation \cite{ADM} which is based on a change of variables (from the
components $g_{0\mu}$ to the lapse and shifts functions) with its linearized
version as it is not clear how linearization of the lapse $N=\left(
-g^{00}\right)  ^{-1/2}$ and shift $N^{i}=-\frac{g^{0i}}{g^{00}}$ can even be
performed (in \cite{ADM-1} the linearized version of GR was discussed using
the metric tensor, not the ADM variables). In addition, the ADM formulation
(the only one in which a restoration of gauge invariance from the complete set
of first class constraints has even been considered before) does not lead to
the expected diffeomorphism invariance as was recently demonstrated in
\cite{Saha} using the method \cite{Novel}, which differs from Castellani's.
The derivation of \cite{Saha} is the most complete one in the literature but
it is not new; the gauge transformations of the ADM variables have been
discussed in part previously in \cite{Berg} and \cite{Castellani}. The gauge
transformations that follow from the ADM formulation can be presented in the
form of diffeomorphism only if \textit{field-dependent} and
\textit{non-covariant} redefinitions of the gauge parameters are performed%
\begin{equation}
\xi_{diff}^{0}=\left(  -g^{00}\right)  ^{1/2}\varepsilon_{ADM}^{\perp
}~,\left.  {}\right.  \left.  {}\right.  \xi_{diff}^{k}=\varepsilon_{ADM}%
^{k}+\frac{g^{0k}}{g^{00}}\left(  -g^{00}\right)  ^{1/2}\varepsilon
_{ADM}^{\bot}~, \label{redef}%
\end{equation}
which, according to \cite{Saha}, \textquotedblleft demonstrate the unity of
the different symmetries involved\textquotedblright. The transformations of
\cite{Saha} are consistent with the transformations obtained in
\cite{Banerjee} using the Lagrangian approach of \cite{Gitman}. However, as we
have already pointed out in \cite{affine-metric}, the field-dependent
redefinition of gauge parameters contradicts to the essence of all known
algorithms for the restoration of gauge invariance, as all of them start from
the assumption that the gauge parameters should be independent of fields (for
details see \cite{affine-metric}). Nevertheless, it would be interesting to
know whether the linearized ADM gravity could produce the linearized version
of the gauge transformations in \cite{Saha} and \cite{Banerjee}, and how
(\ref{redef}) would look in the linearized case.

Almost immediately after the appearance of \cite{Saha} with ADM gravity, the
restoration of gauge transformations in GR was considered by Samanta
\cite{Samanta} using the Lagrangian approach of \cite{Gitman}. This method
indeed leads to the diffeomorphism invariance when the metric or metric and
affine connection (for the first-order formulation) are used as the
fundamental variables and no non-canonical transformations are performed. We
also confirmed in \cite{affine-metric} that in the Hamiltonian formulation of
the first-order, affine-metric, GR the diffeomorphism symmetry is generated by
the first class constraints.

We now would like to note, that despite the similarities of the Hamiltonian
formulations of linearized and full GR, there are fundamental differences
between them. In particular, the Hamiltonian of full GR is proportional to
secondary constraints \cite{Dirac, myths, PLA}, the algebra of the secondary
constraints has field-dependent structure functions, and the equations of
motion are invariant under diffeomorphism transformation only
\textquotedblleft on-shell\textquotedblright\ \cite{myths}. In the linearized
case the Hamiltonian is not a constraint, the algebra of constraints does not
depend on fields, and the equations of motion are invariant exactly under
(\ref{eq.23}). These differences might be crucial when one tries to project
results obtained for linearized gravity to full GR. Consequently, using the
Hamiltonian formulation of linearized GR as a guide has to be done cautiously.

We would like to conclude our paper with words of Carmeli \cite{Carmeli}:
\textquotedblleft our experience shows that solutions of the linearized
equations may bear little or no relation to solutions of the rigorous
equations\textquotedblright; and later: \textquotedblleft One should therefore
in no way consider the linearized theory as being a substitute to the full
theory\textquotedblright.

The distinction between full and linearized GR is under our current
investigation and the results will be reported elsewhere.

\section{Acknowledgements}

We are grateful to A.M. Frolov, D.G.C. McKeon and S.R. Valluri for helpful
discussions. The research was partially supported by the Huron University
College Faculty of Arts and Social Science Research Grant Fund.

\end{document}